# MICRO-BALL LENS ARRAY FABRICATION IN PHOTORESIST USING PTFE HYDROPHOBIC EFFECT

Ruey Fang Shyu[1], Hsiharng Yang[2, 3], Wen-Ren Tsai[2] and Jhy-Cherng Tsai[4]

[1]Department of Mechanical Manufacturing Engineering, National Formosa University, Yunlin, Taiwan 632
[2]Institute of Precision Engineering, National Chung Hsing University, Taichung, Taiwan 402
[3]Center of Nanoscience and Nanotechnology, National Chung Hsing University, Taichung, Taiwan 402
[4]Department of Mechanical Engineering, National Chung Hsing University, Taichung, Taiwan 402

**ABSTRACT**

This paper presents a simple method to fabricate micro-ball lens and its array. The key technology is to use the hydrophobic characteristics of polyterafluoroethylene (PTFE) substrate. High contact angle between melted photoresist pattern and PTFE can generate micro-ball lens and its array. PTFE thin film was spun onto a silicon wafer and dried in oven. Photoresist AZ4620 was used to pattern micro-columns with different diameters 60, 70 and 80 μm. A thermal reflow process then was applied to melt these micro-column patterns resulted in micro-ball lens array. The achieved micro-ball lens array with diameter 98 μm was fabricated using 80 μm in diameter patterns. This method provides a simple fabrication process and low material cost.

**1. INTRODUCTION**

Miniaturized optical devices are very interested in portable instruments and optical communications. A micro-ball lens can be used to obtain more efficient light coupling from a laser diode (LD) into a single-mode fiber (SMF). This is necessary for low-loss and high-bit-rate communication systems. The primary challenge for these optical devices results from the extremely stringent submicron alignment tolerance required for high-efficiency coupling. A ball lens can achieve low-cost coupling, but the coupling efficiency of the general ball lens is limited to around 10％. This poor performance is caused by the large spherical aberration size. Increasing ball lens mode converter performance through a more reliable design has been discussed by many researchers using theoretical or experimental methods. Correctly predicting the refractions from a ball lens into a SMF [1] and accurately calculating the coupling efficiency from a LD into an optical fiber through a ball lens using the exact solution to Maxwell's equations for the beam scattering from a sphere has been accomplished [2]. Maxwell's equations compared to a conventional method using geometrical and diffraction analysis have also revealed that the optimum coupling efficiency depends on sufficiently fine discretization of the distances between the source and fiber from the ball lens [3]. An efficient coupling scheme between a SMF and a silica guide using a ball lens has been demonstrated using theoretical analysis and experiments [4]. Micro-ball lens provides high coupling efficiency up to 63% in fiber transmission [5]. The coupling solution for obtaining a very good coupling efficiency with large separations between the optical system elements was produced.

Other ball lens integrated systems such as the wavelength division multiplexing (WDM) transmission system, provide independent channels over a single optical fiber with plural signals of different wavelengths have some advantages such as, cost transmission line and flexible system design. The 1.2/1.3 $\mu$ m band single mode fiber WDM was developed with various factors using ball lens collimators in the optical coupling system. They achieved all design objects including high performance, high stability and versatility [6]. A new integrated optical component, named the beam-splitting ball lens, can be used in conjunction with polymer fiber image guides to split and combine two dimensional arrayed optical data imaged patterns. This method will help ease the design concerns for future 2D array-based large-bandwidth board and backplane level optical interconnections [7]. In hybrid free-space MOEMS (micro-opto-electro-mechanical systems) chip integration, a silicon micromachined submount is now under rapid development. The submount is designed to accommodate various free-space MOEMS chips and reduce the packaging cost using minimal active optical alignment. A micro-ball lens can be applied in MOEMS. The micro-optical switch (an indispensable element for optical communication systems) is representative of these devices. It offers an all-optical network to increase the data exchange speed and light propagation performance. A micro mirror switch packaged with MOEMS device, including four V-grooves for optical fibers and micropits for micro-ball lenses, was assembled using the standard integrated circuit technique. The experimental test showed that the micro mirror switch vibration was up to 100 g's and the frequency was from 200 Hz to 10 kHz for over 24 hours [8]. A batch assembly micro-ball lens array with a diameter lower 50 $\mu$ m was also developed using optical coupling in free-space communications [9, 10]. This device will achieve the lens miniaturization goal.

Many articles related to micro-ball lens applications have appeared in the technical literature, however, papers regarding micro-ball lens fabrication methods different from the general microlens array are rare. The heat reflow technique is commonly used in the refractive microlens array to fabricate a planar microlens array by melting cylindrical resist posts onto a substrate [11 - 13]. The material surface is melted once the heating temperature is above the material glass temperature. Surface tension effects on the melted material surface result in a spherical profile. The fabrication method based on the heat reflow of two polymer layers to fabricate an integrated micro-ball lens array onto a planar substrate was studied [5]. Long processing time in the thermal reflow was the critical issue. The other melting polymer method by using oil-bath heating was presented [14]. A very challenge on making PTFE patterns for micro-ball lens was found. They contribute to investigate by using a simple process for micro-ball lens fabrication.

## 2. MICRO-BALL LENS FORMATION PRINCIPLE

The micro-ball lens formation principle is similar to the general wetting surface situation. Once a drop contacts the substrate, the drop will spread onto the substrate when its surface energy reaches the $S_{sa} > S_{ls} + S_{al}$ inequality. Where $S_{sa}$ is the surface energy of the solid. And $S_{ls}$ and $S_{al}$ represent the solid-liquid and liquid surface energy, respectively. If the surface energy has a relationship of $S_{sa} < S_{ls} + S_{al}$, the drop will form a ball shape because of the decreasing drop contact area with the substrate. Air tends to minimize the surface energy.

When the patterned polymer is heated above its glass transition temperature, the melting polymer surface will change into a spherical profile as illustrated in Figure 1. According to Young's modulus, the equation for a liquid drop on a solid surface can be shown as $S_{sa} = S_{ls} + S_{al} \cos\theta$. Where $\theta$ is the equilibrium contact angle. According to the equilibrium for the above equation, the contact angle of interface will gradually increase when solid-liquid interface energy increases. A drop will form a ball when the contact angle becomes greater than a critical value. Some surfaces have liquid contact angles as high as 150° or even 180°. On these surfaces, a droplets simple rest on the surface, without actually wetting to any significant extent. These surfaces are termed super hydrophobic and can be obtained on fluorinated surfaces (Teflon-like coatings) that have been appropriately patterned. The contact angle thus directly provides information on the interaction energy between the surface and the liquid. The PTFE substrate provides the key role to affect the contact angle of the liquid such as the melted photoresist. Surface roughness of PTFE, micro-ball lens geometric related to photoresist patterns, and fabrication processing parameters were studied.

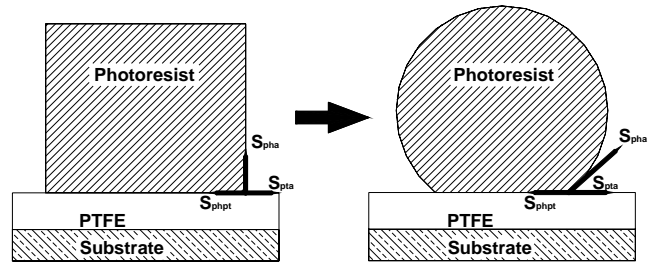

Figure 1. Illustration of a micro-ball lens formation onto PTFE coating substrate in the thermal reflow process.

## 3. THEPRETICAL MODELS

The geometrical dimension of a micro-ball lens before and after heat reflow is shown in Figure 2. The photoresist pattern change into a ball was the consideration in this study. The mass conservation law was used to develop the mathematical model for designing and fabricating a micro-ball lens. Before the heat reflow process, the photoresist pattern volume is expressed in Equation (1).

$$V_{column} = \frac{\pi d_b^2 t_b}{4} \qquad (1)$$

Where $d_b$ is the diameter of the photoresist pattern and $t_b$ is its thickness. A contact area is necessary to stabilize and fix the ball lens on the substrate. Therefore, the ball lens volume actually results from the cap and the spherical area cut away by the contact area. The accurate volume calculation is shown in Equation (2).

$$\begin{aligned}V_{ball-lens} &= \frac{1}{6}\pi D_b^3 - \frac{1}{24}\pi D_b^3 (2+\cos\theta)(1-\cos\theta)^2 \\ &= \frac{1}{6}\pi D_b^3 [1 - \frac{1}{4}(2+\cos\theta)(1-\cos\theta)^2]\end{aligned} \qquad (2)$$

Where $\theta$ is the contact angle between the all lens and substrate. According to the mass conservation law, Equation (1) is equal to Equation (2). It results in Equation (3).

$$\frac{\pi d_b^2 t_b}{4} = \frac{1}{6}\pi D_b^3 [1 - \frac{1}{4}(2+\cos\theta)(1-\cos\theta)^2] \qquad (3)$$

Where $d_b$ can be obtained from the mask design and $t_b$ is determined from the spin coating experiments as shown in Figure 3. The primitive experiment showing the contact angle 116° was determined Equation (4).

$$\theta = 90° + \tan^{-1} \frac{2h - 2R}{\sqrt{8Rh - 4h^2}} \quad (4)$$

Where R is the radius and h is the sag height of the ball lens. Further calculations may result the spin speed related to ball lens diameter and sag height respect to various sizes of photoresist patterns as shown in Figure 4 and 5.

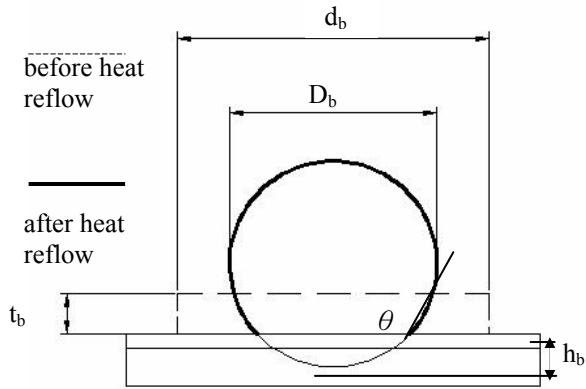

Figure 2. Schematic diagram of the volumetric variation and related geometry of a micro-ball lens before and after thermal reflow process.

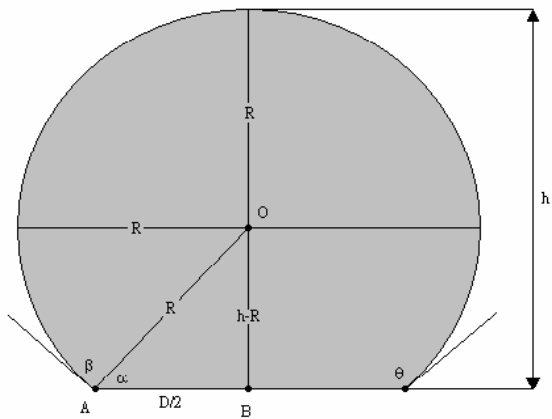

Figure 3. Geometrical dimensions of a micro-ball lens.

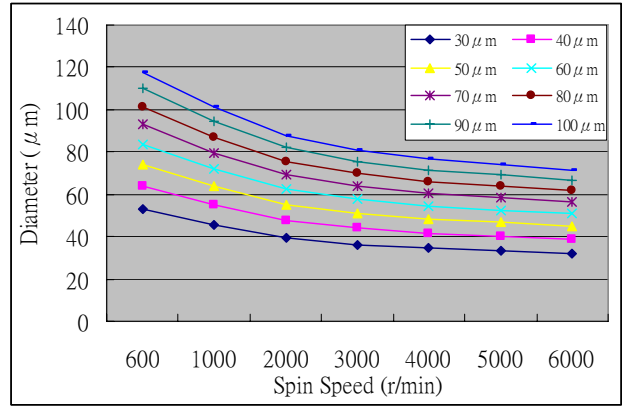

Figure 4. The relationship between spin speed and micro-ball lens diameter with respect to various sizes of photoresist patterns.

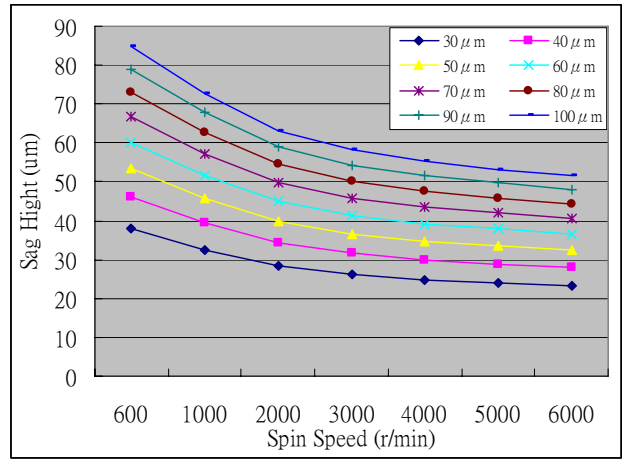

Figure 5. The relationship between spin speed and lens sag height with respect to various sizes of photoresist patterns.

## 4. EXPERIMENTAL METHODS
### 4.1 Fabrication process
In this experiment, the pattern on the mask was designed to various sizes in array with diameters 60, 70, and 80 μm. A schematic diagram of the entire process of micro-ball lens array fabrication process is shown in Figure 6. Silicon wafers were used as the substrate. The wafers were firstly cleaned and dehumidified first in an oven at a temperature of 150℃ for 30 minutes. The wafers were then coated with a liquid PTFE layer (Land Holder Trading Co., Ltd. ALGOFLOND 60/A) using a spin coater. The physical parameters of liquid PTFE are listed in Table 1. The coating process of PTFE is listed in Table 2. The spin condition was 1000 rpm for 10 seconds and 2000 rpm for 40 seconds. Three baking steps were used. First step was to remove most solvent at temperature 110 °C for 10 minutes in an oven. Second step used a baking temperature 165°C for 10 minutes to remove remaining solvent. Third step was sued to increase its surface adhesion onto the substrate. The baking temperature was 260 °C for 15 minutes. Exceeding baking time to 30 minutes for both step 2 and 3 may result in PTFE layer peel-off from the substrate.

Desired patterns were transferred from the designed mask in the lithographic process. In this experiment, a plastic mask was fabricated using a laser writing onto a PET (Polyethylene terephthalate) used for PCBs (print circuit boards). The design pattern on the plastic mask is illustrated in Figure 7. Each pattern was 80 μm in diameter. The geometry was measured by an optical imagery tool (Optimas). A positive photoresist AZ4620 was patterned onto the substrate in lithography process as the result illustrated in Figure 6(b). The sample was placed into the oven and heated from room temperature (25°C) to 120 °C in 10 minutes illustrated in Figure 6(c). Rapid heating may result in the sample cracking due to thermal expansion. Then heating up to the photoresist glass transition temperature 160 °C for 15 minutes, a thermal reflow process to melt the photoresist is required. The sample was still hold in the oven to cool down to room temperature naturally. Micro-ball lens array was formed onto the substrate coated with PTFE as indicated in Figure 6(d).

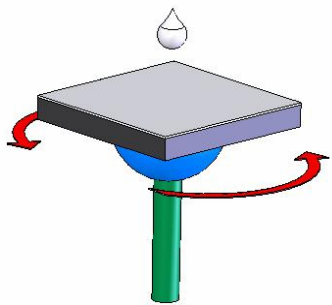

(a)

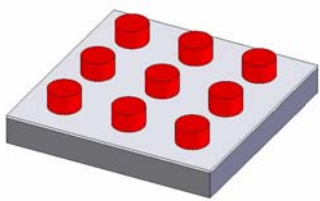

(b)

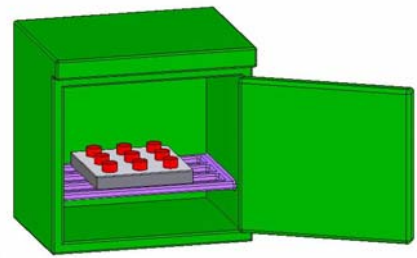

(c)

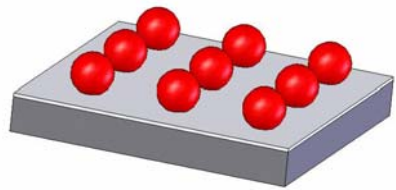

(d)

Figure 6. Schematic of micro-ball lens array fabrication; (a) PTFE coating, (b) pattern transfer by lithography process, (c) thermal reflow in an oven, and (d) micro-ball lens array formation.

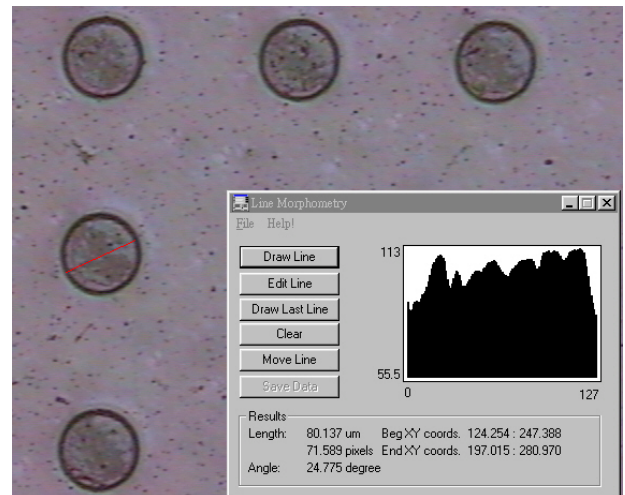

Figure 7. Pattern design on the mask.

Table 1. Physical parameters of liquid PTFE.

| Melt Point °C (°F) | 340 (644) |
|---|---|
| PTFE content(%) | 60 |
| Non Ionic surfactant (on the mixture)(%) | 3 |
| Non Ionic surfactant (on the solid)(%) | 6 |
| pH | 9 |
| Specific Gravity at 20°C | 1.52 |
| Conductivity( $\mu$ s/cm) | 700 |

| Avg. part. size($\mu$m) | 0.24 |
|---|---|
| Brookfield viscosity at 35°C(mPa.s) | 20 |

Table 2. PTFE coating process.

| Process Step | Parameters | Equipment |
|---|---|---|
| Clean | 1. $H_2SO_4$ : $H_2O_2$ = 3:1<br>2. DI Water for 5 min.<br>3. $N_2$ dry<br>4. Bake 120°C | Chemical Hood<br>Ultrasonic cleaning<br><br>Oven |
| PTFE coating | 1. Spread :<br>    1000 rpm for 10 s<br>2. Spin :<br>    2000rpm for 40 s | Spin coater<br>Liquid PTFE |
| Bake | 1. 10 min. at 110°C<br>2. 10 min. at 165°C<br>3. 10~15 min. at 260°C | Hot plate |

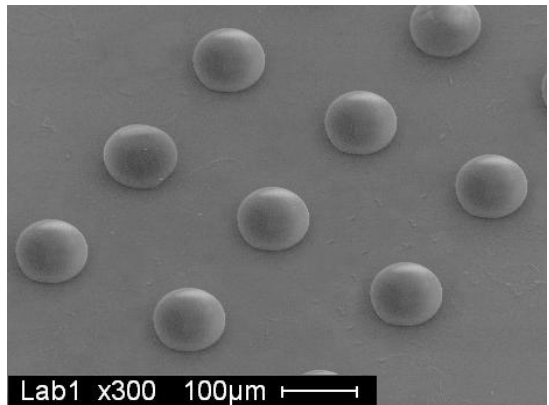

(a)

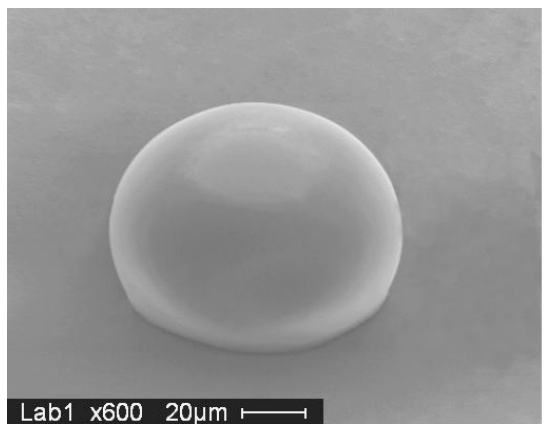

(b)

Figure 8. Experimental results of micro-ball lens; (a) SEM micrograph of the micro-ball lens array and (b) an individual micro-ball lens.

## 4. DISCUSSION

### 4.1 Surface quality

Surface roughness measurement of the micro-ball lens performed using a NanoFocus μScan 3D laser profilometer is shown in 9. The measured length was 20 μm. The average surface roughness Ra = 7.6 nm for 10 measurements. High surface quality was achieved by using the thermal reflow process for the micro-ball lens fabrication.

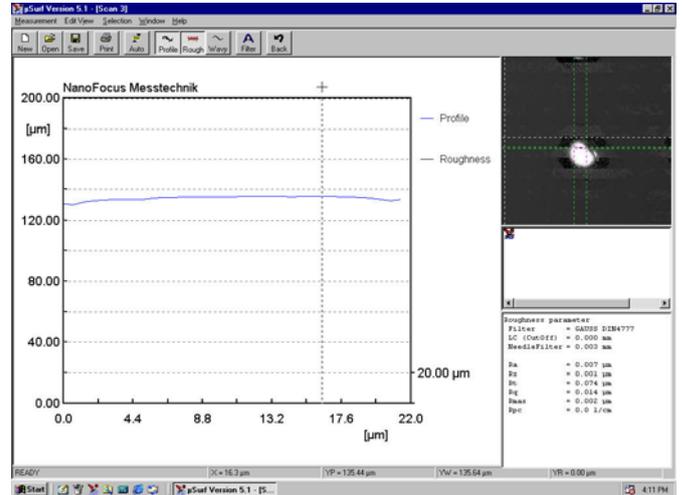

Figure 9. Surface roughness measurement by using a NanoFocus μScan 3D laser profilometer.

### 4.2 Microlens geometrical verification

The visualized micro-ball lens was shown from the SEM micrograph. So far, there is no solution to direct detect a standing micro-ball lens in three dimensions. Top portion of the micro-ball lens was measured by using a stylus surface profiler as shown in Figure 10. A theoretical curve of a micro-ball lens profile is sketched for comparing their deviations. Figure 10 is the micro-ball lens formed by using pattern diameter 80 μm. The errors of the micro-ball lens in diameter and height are 3.3 and 4.0 %, respectively. Table 3 lists the comparison of different pattern diameters in experimental and theoretical values.

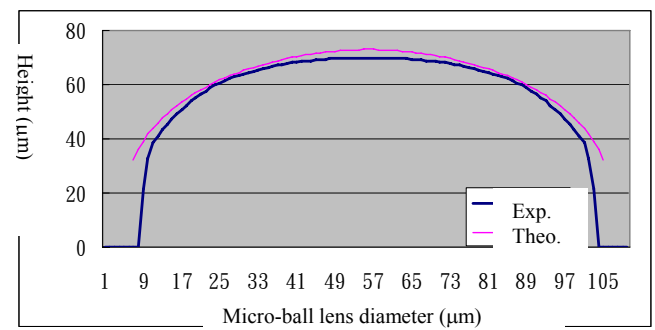

Figure 10. Micro-ball lens geometrical profile.

Table 3. Comparison of experimental and theoretical geometrical profile of micro-ball lenses.

| Pattern diameter<br>Micro-ball lens | 80μm | 70μm | 60μm |
|---|---|---|---|
| Avg. Exp. Dia. (μm) | 98.20 | 89.21 | 82.32 |
| Theoretical Dia, (μm) | 101.59 | 92.94 | 86.39 |
| Error in Dia. (%) | 3.3 | 4.0 | 4.7 |
| Avg. experimental height (μm) | 70.18 | 62.18 | 56.95 |
| Theoretical height (μm) | 73.07 | 66.80 | 60.00 |
| Error in height (%) | 4.0 | 6.9 | 5.1 |

## 4. CONCLUSION

A simple process for fabricating micro-ball lens and its array onto PTFE coating substrate was completed. The theoretical achievable micro-ball lens diameters are ranged from 30 to 110 μm. The experimental result showed that the micro-ball lens 60, 70 and 80 μm in diameter less than 5% uncertainty were achieved. The micro-ball lens array can be provided for high dense optical communication systems and other optical coupling applications.

## 6. ACKNOWLEDGEMENT

This work was supported by the National Science Council (series no. NSC 94-2212-E-150 -016) of Taiwan, R.O.C.